\documentclass[12pt]{article}
\usepackage{amsfonts,epsf}
\def\d{{\rm d}}
\def\mi{{\rm i}}

\def\g{\mathop{\gamma}\nolimits}
\def\G{\mathop{\Gamma}\nolimits}
\def\Re{\mathop{\rm Re\,}\nolimits}

\def\e{\mathop{\rm e}\nolimits}
\def\Res{\mathop{\rm Res}\nolimits}
\def\hf{{\textstyle{1 \over 2}}}
\def\qt{{\textstyle{1 \over 4}}}
\def\od{{\textstyle{1 \over 2}+{1 \over N}}}

\def\defi{\stackrel{\rm def}{=}}
\def\si{\!\!\! &}
\def\sf{& \!\!\!}
\def\beq{\begin{equation}}
\def\eeq{\end{equation}}
\def\bea{\begin{eqnarray}}
\def\eea{\end{eqnarray}}

\title{From exact-WKB towards\\
singular quantum perturbation theory}
   
\author{{\bf Andr\'e Voros}\\
\\
CEA, Service de Physique Th\'eorique de Saclay\\
CNRS URA 2306\\
F-91191 Gif-sur-Yvette CEDEX, France\\
{ E-mail : {\tt voros@spht.saclay.cea.fr}}\\
(also at: Institut de Math\'ematiques de Jussieu--Chevaleret, CNRS UMR 7586\\ Universit\'e Paris 7, F-75251 Paris Cedex 05, France)}

\begin{document}
\maketitle

\begin{abstract}
We use exact WKB analysis to derive some concrete formulae 
in singular quantum perturbation theory, 
for Schr\"odinger eigenvalue problems on the real line with polynomial potentials 
of the form $(q^M+g q^N)$, where $N>M>0$ even, and $g>0$. Mainly,
we establish the $g \to 0$ limiting forms of global spectral functions
such as the zeta-regularized determinants and some spectral zeta functions.
\end{abstract}

\section{Introduction}

The purpose of this work is to set up a path to obtain precise statements
of a quantum perturbative nature with the help of exact WKB analysis.
The RIMS has always played a major and pioneering role in 
the inception and growth of exact asymptotic analysis, and earlier,
in the development of some of its fundamental tools
(such as hyperfunctions and holomorphic microlocal analysis).
This influence is testified by the Proceedings volume of a recent Kyoto 
conference, which offers a very complete view of the subject \cite{HKT}. 
It is therefore a great honor and a proper tribute 
to RIMS to write about exact WKB analysis in this anniversary issue.
\medskip

A prototype problem in quantum perturbation theory is 
the quartic anharmonic oscillator,
\beq
\label{QAO}
\Bigl( -{\d^2 \over \d q^2}+ q^2 + g q^4 - E \Bigr) \Psi (q) = 0, \qquad 
q \in {\mathbb R},\ g \ge 0 .
\eeq
This problem has a purely discrete eigenvalue spectrum $\{ E_k(g) \}$
for all $g \ge 0$. 
A typical task in (Rayleigh--Schr\"odinger) perturbation theory is 
to compute individual eigenvalues $E_k(g)$ (or their eigenfunctions)
as formal power series of the coupling constant $g$ \cite{BW}. 
This is practically important when the unperturbed ($g=0$) problem 
is exactly solvable, here a harmonic oscillator;
a major drawback is however that the coupling term has the higher degree, 
hence the formalism is singular.
Thus, the perturbation series converges for no $g \ne 0$;
it only gives an asymptotic expansion for $g \to 0$, 
moreover non-uniform in the quantum number $k$.

As our theoretical discussion can readily include all binomials potentials, 
we will actually study the more general Schr\"odinger equation
\beq
\label{GAO}
\Bigl( -{\d^2 \over \d q^2}+ U_g(q) - E \Bigr) \Psi (q) = 0, \qquad 
U_g(q) \defi q^M + g q^N , \quad q \in {\mathbb R}, 
\ N>M \ge 2 \mbox{ even}, \ g \ge 0 ;
\eeq
we keep $E_k(g)$ as a generic notation for the eigenvalues of this problem 
($(N,M)$-dependences now being implied).

(Exact WKB formalisms accommodate non-even potentials as well \cite{DP,VQ};
for instance, eq.(\ref{GAO}) could be considered with odd $N$ or $M$ 
but on the half-line $[0,+\infty)$ \cite{VQ}; 
however, this extension is not essential here while it does complicate
the classification when $M=1$, so we omit it in the present work.)

A very basic fact ({\sl Symanzik scaling\/} property) is that
a simple coordinate dilation, $q \mapsto g^{-1/(N+2)} q$, 
establishes a unitary equivalence between the two Schr\"odinger operators
\bea
\label{SYZ}
v^{2/(M+2)} (- \d^2 / \d q^2 + U_g(q)) \quad &\mbox{and}& \quad
\hat H_v \defi - \d^2 / \d q^2 + V_v(q), \\
\mbox{where} \qquad v \equiv g^{-(M+2)/(N+2)} \quad &\mbox{and}& \quad 
V_v(q) \defi q^N + v q^M .
\eea
Thus, eq.(\ref{GAO}) is equivalent to
\beq
\label{NMO}
\Bigl(- {\d^2 \over \d q^2}+ q^N + v q^M + \lambda \Bigr) \Psi (q) = 0, \qquad 
v \equiv g^{-(M+2)/(N+2)}, \ \lambda \equiv -v^{2/(M+2)} E .
\eeq

In this transformed Schr\"odinger equation, 
the interaction term is now $vq^M$ and has the lower degree,
so that $v$ can act as a regular deformation parameter;
the former perturbative regime $g \to 0$
translates as the asymptotic $v \to +\infty$ regime. However, 
at no finite $v$ is the problem (\ref{NMO}) solvable in any traditional sense,
and this has severely limited practical uses of this reparametrization. 
On the other hand, this deformation can be fully studied by exact WKB analysis, 
which now handles general (1D) polynomial potentials.
One earlier detailed study of this sort is based on resurgence theory \cite{DP}.
Another such path from exact WKB to perturbation theory lies in
proving the Zinn-Justin conjectures about multi-instantons \cite{DDP,Z}.
Here, continuing a different type of study initiated in \cite{VQ} (Secs.~3--4)
(within an exact WKB framework built upon Sibuya's formalism \cite{S}), we seek 
to specify how the {\sl spectral determinants\/} themselves (and related spectral functions) 
{\sl asymptotically depend on the coupling parameter\/} $v \to +\infty$ (or $g \to 0$). 
Spectral functions being symmetric functions of all eigenvalues $E_k(g)$ together,
the non-uniformity in $k$ of perturbative approximations must show up somehow,
and the $g \to 0$ behavior of such objects might not be obviously traceable to
existing (fixed-$k$) perturbative results.

As a very concrete example, we may ask: how do the spectral zeta functions
$Z_g (s) = \sum_{k=0}^\infty E_k(g)^{-s}$ precisely behave for $g \to 0$?
Specially at $s=1$ when $M=2$: then, that series converges $\forall g>0$ but
term by term it becomes the divergent (odd) harmonic series $\sum_k (2k+1)^{-1}$ at $g=0$.
The latter admits one fundamental regularization by means of a sharp summation cutoff $K$:
\beq
\label{GM}
\sum_{k=0}^{K-1} {1 \over 2k+1} \sim \hf (\log K + \g + 2 \log 2),
\qquad K \to +\infty ,
\eeq
which is (in just a slight variant form) the basic definition of
Euler's constant $\g$. Now, the eigenvalues themselves obey
$E_k(g) \sim 2k+1$ for $g \to 0$ ($k$ fixed) by perturbation theory,
but $E_k(g) \propto [g(2k+1)^N]^{2/(N+2)}$ for $k \to +\infty$ ($g$ fixed)
by the asymptotic Bohr--Sommerfeld condition, the crossover zone
being roughly located by $2K_g +1 \propto g^{-2/(N-2)}$ \cite{HMM}.
Hence the series for $Z_g (1)$ is another natural, ``soft" regularization
of the odd harmonic series, beginning to act around $k \propto K_g$.
Substituting $K=K_g$ into eq.(\ref{GM}), we are led to expect
$Z_g (1) \sim -{1 \over N-2} \log g + C_N $.
Such an intuitive approach may work for the logarithmic slope,
but not for identifying the additive constant $C_N$.
By contrast, exact WKB analysis can yield 
a {\sl precise asymptotic prediction\/} for this (and other) zeta-values:
see our final formula (\ref{RF}).
\medskip

The outline of the paper is as follows. 
Sec.~2 recalls essential prerequisites and definitions
for the exact WKB approach to be used here. 
Sec.~3 presents the asymptotic problem and its
conceptual resolution by exact WKB theory as eq.(\ref{CMP}). 
Sec.~4 performs the key computational steps: a class of specific 
{\sl improper action integrals\/} $\int_0^{+\infty} \Pi(q) \, \d q$
(where $\Pi(q)$ are essentially classical momentum functions,
and the integrals are primitively very divergent) are explicitly evaluated. 
Finally, Sec.~5 processes all intermediate calculations into concrete formulae,
mainly eqs.(\ref{AS}),(\ref{ASF}).

\section{Some notions from exact WKB theory}

We recall the essential facts and notations to be used later concerning
the exact WKB treatment of Schr\"odinger operators on $L^2({\mathbb R})$, 
of the form 
$\hat H \defi - \d^2 / \d q^2 + V(q)$, with a polynomial potential 
$V(q)= +q^N + \mbox{[lower-degree terms]}$, here taken {\sl real and even\/}.
Details and justifications have to be omitted 
(cf. \cite{VQ} Sec.~1, and references therein).
Such operators are self-adjoint, have compact resolvents and commute with
the parity operator $(q \mapsto -q)$. A frequently needed quantity 
(which we call the {\sl order\/} of the problem) is
\beq
\label{ORD}
\mu(N) \defi {1 \over 2} + {1 \over N} \ .
\eeq

As standard notations, we will also use $\psi(z) \equiv [ \G' / \G ](z)$, 
and $\g =$ Euler's constant.

\subsection {Improper action integral, and residue function}

Important quantities enter at the classical dynamical level around 
the (complexified) momentum function, 
\begin{equation}
\label{PI}
\Pi_\lambda(q) \defi (V(q)+\lambda) ^{1/2}, 
\end{equation}
where the constant $(-\lambda)$ stands for the classical energy and,
say, $\lambda > - \inf \ V$ (initially). 
Next, {\sl improper action integrals\/} over semi-infinite paths prove 
very useful: $\int_q^{+\infty} \Pi_\lambda(q') \,\d q'$ 
(primitively divergent) is very naturally redefined as 
the analytical continuation to $s=0$, {\sl when this is finite\/}, of
\beq
\label{IQ}
I_q(s,\lambda) \defi
\int_q^{+\infty} (V(q')+\lambda)^{-s+1/2} \,\d q' \qquad 
(\mbox{convergent for } \Re(s) >\mu(N)) .
\eeq
Now, the $s$-plane singularities of $I_q(s,\lambda)$ entirely come from
the large-$q$ behavior of the integrand.
Specifically, the $q \to +\infty$ expansion (explicitly computable order by order) 
\begin{equation}
\label{BET}
(V(q) + \lambda)^{-s+1/2} \sim 
\sum_\rho \beta_\rho (s) q^{\rho-Ns} \qquad \textstyle
(\rho=N/2,\ N/2 \, -1, \cdots) 
\end{equation}
implies the singular decomposition
\begin{equation}
\label{IA}
I_q(s,\lambda) \sim 
-\sum_\rho \beta_\rho(s) {q^{\rho+1-Ns} \over \rho+1-Ns} \, ;
\end{equation}
hence at $s=0$, $I_q(s,\lambda)$ has at most a simple pole, 
generated by the $\rho=-1$ term (if any):
\beq
\label{RES}
\Res_{s=0} I_q(s,\lambda) = \beta_{-1}(0)/N ,
\eeq
a value actually independent of $\lambda$ (save when $N=2$) and of $q$.
\medskip

A central distinction sets in at this point: 
if the ``residue function" $\beta_{-1} (s) \equiv 0$, 
the Schr\"odinger problem $(\hat H +\lambda) \Psi =0$ will behave more simply
(``normal" type, {\bf N}); 
otherwise, ``anomaly" corrections will enter (type {\bf A}). 
Wholly generic polynomials $(V(q) + \lambda)$ are of type {\bf A}; 
still, in a sense, ``a majority" of them have $\beta_{-1} (s) \equiv 0$: 
among the even ones, already all those having a {\sl degree $N$ multiple of\/} 4
(and, among the non-even ones, all those of {\sl odd degree\/}).

Thus, for the {\bf N} type, $\int_q^{+\infty} \Pi_\lambda(q') \,\d q'$ 
can be readily defined as 
the analytical continuation of $I_q(s,\lambda)$ to the (regular) point $s=0$; 
whereas for the general ({\bf A}) type, the best specification is not 
the bare finite part of $I_q(s,\lambda)$ at the pole $s=0$ 
(denoted ${\rm FP}_{s=0}I_q(s,\lambda)$), but rather (\cite{VQ}, eq.(32))
\beq
\label{IFP}
\int_q^{+\infty} \!\! \Pi_\lambda(q') \,\d q' \defi 
{\rm FP}_{s=0} I_q(s,\lambda) + 2(1-\log 2) \,\beta_{-1}(0) / N 
\eeq
in order to preserve the basic identities (\ref{ID}) below. Additivity is also maintained:
\beq
\label{ADD}
\int_q^{+\infty} \!\! \Pi_\lambda(q') \,\d q' =
\int_q^{q''} \!\! \Pi_\lambda(q') \,\d q' + \int_{q''}^{+\infty} \!\! \Pi_\lambda(q') \,\d q'
\qquad \mbox{for all $q$, $q''$ finite}
\eeq
(because a finite integral $\int_q^{q''} (V(q')+\lambda)^{-s+1/2} \,\d q'$ is entire in $s$).
\medskip

Remarks: (i) $\int_q^{+\infty} \Pi_\lambda(q') \,\d q'$ is an ``Agmon distance 
from $q$ to $+\infty$", suitably renormalized;
(ii) this procedure is a classical counterpart to zeta-regularization at the quantum level;
(iii) just as the extra term in eq.(\ref{IFP}), 
all anomaly terms here will simply be proportional
to the value $\beta_{-1}(0)$, but more general forms occur elsewhere \cite{VQ}.

\subsection {Spectral functions}
 
An operator $\hat H$ as above has a purely discrete real spectrum 
$\{ \lambda_0 < \lambda_1 < \lambda_2 < \cdots \}$,
($\lambda_k \uparrow +\infty$), where even (resp. odd) $k$ correspond to 
eigenfunctions of even (resp. odd) parity. 
Parity-split spectral zeta functions ({\sl \`a la\/} Hurwitz) can be defined as
\beq
Z^\pm (s,\lambda) \defi 
\sum_{k \ {\rm even \atop odd}} (\lambda_k + \lambda)^{-s}
\qquad \mbox{for } \Re s>\mu(N)
\eeq
and, say, $\lambda > -\lambda_0$; many results however take a sharper form upon
a {\sl skew\/} versus a {\sl full\/} zeta function, respectively defined as
\beq
Z^{\rm P} \equiv Z^+ - Z^-, \qquad Z \equiv Z^+ + Z^- .
\eeq
Spectral determinants $ D^\pm(\lambda) \equiv {\det}^\pm (\hat H + \lambda) $
are defined by zeta-regularization, as
\beq
\label{DZR}
D^\pm(\lambda) \defi \exp [-\partial_s Z^\pm (s,\lambda)]_{s=0}
\qquad (\mbox{and } D^{\rm P} \equiv D^+/D^-, \quad D \equiv D^+ D^-),
\eeq
where $s=0$ is reached by analytical continuation from $\{ \Re s>\mu(N) \}$.
More constructive specifications are

\noindent - their Weierstrass infinite products 
(written for $\mu(N) < 2$, which is true here):
\bea
\label{WPH}
D^\pm(\lambda) \si \equiv \sf D^\pm(0) \e^{ {\rm FP}_{s=1}Z^\pm(s,0) \,\lambda }
\prod_{k \ {\rm even \atop odd}} 
( 1 + \lambda / \lambda_k ) \e^{-\lambda / \lambda_k} , \\
\label{WPR}
\si \equiv \sf D^\pm(0) \prod_{k \ {\rm even \atop odd}} 
\!(1 + \lambda / \lambda_k) \qquad \mbox{when $\mu(N) < 1$, i.e., $N>2$}
\eea
and likewise for $D$, $D^{\rm P}$; this shows that the determinants continue 
to {\sl entire\/} functions (of order $\mu(N)$) in the variable $\lambda$
(except for $D^{\rm P}$, {\sl meromorphic\/});

\noindent - the basic identities of the exact-WKB method: let $\Psi_\lambda(q)$ 
be the {\sl canonical recessive\/} solution of the differential equation, 
specified through its $q \to +\infty$ {\sl asymptotic form\/}
\beq
\label{WKB}
\Psi_\lambda (q) \sim \Pi_\lambda(q)^{-1/2} \e^{ \int_q^{+\infty} \Pi_\lambda(q') \d q'} ,
\eeq
where $\int_q^{+\infty} \Pi_\lambda(q') \, \d q'$ is
{\sl fixed according to eq.(\ref{IFP})\/}; then, under that precise normalization,
\beq
\label{ID}
D^-(\lambda ) \equiv \Psi_\lambda (0), \qquad D^+(\lambda ) \equiv -\Psi'_\lambda (0) ,
\eeq
(also valid for a rescaled potential, i.e., $V(q)= u q^N+ \cdots$, with $u>0$).
Remark: the solutions obeying (\ref{WKB}) are close to 
Sibuya's subdominant solutions \cite{S}, 
but the two normalizations fully coincide only when the type is {\bf N}.

\medskip

Finally, we will need the transformation rules for spectral functions
under a {\sl global spectral dilation\/}
($ \lambda_k \mapsto r \lambda_k $, $r = {\rm cst.} >0$). Obviously,
\beq
Z^\pm (s,\lambda) \mapsto r^{-s} Z^\pm (s, \lambda / r)
\qquad \mbox{for } \Re s>\mu(N)
\eeq
(and likewise for $Z$, $Z^{\rm P}$);
hence upon continuation to $s=0$, and applying eq.(\ref{DZR}),
\beq
\label{SC}
D(\lambda) \mapsto r^{Z(0, \lambda / r)} D(\lambda / r) ,
\qquad D^{\rm P}(\lambda) \mapsto
r^{Z^{\rm P}(0, \lambda / r)} D^{\rm P}(\lambda / r)
\eeq
where, moreover, (\cite{VQ}, eqs.(27),(37))
\beq 
Z(0, \lambda) \equiv -2 \beta_{-1} (0) / N , 
\qquad \qquad Z^{\rm P}(0, \lambda) \equiv 1/2 .
\eeq

\section{The asymptotic $v \to +\infty$ problem}

We now return to the Schr\"odinger operator 
$\hat H(v) = - \d^2 / \d q^2 + q^N + v q^M$, as in eq.(\ref{NMO}) 
($N>M \ge 2$ both even, $v>0$). 
We will find the asymptotic behaviors of its spectral determinants
in the regime of singular perturbation theory for eq.(\ref{QAO}):
\beq
D^\pm (\lambda,v) \equiv {\det}^\pm (\hat H(v) + \lambda) \quad (\lambda>0),
\qquad v \equiv g^{-(M+2)/(N+2)} \to +\infty .
\eeq

To lowest-order in the $g \to 0$ perturbation theory,
the {\sl individual\/} eigenvalues $\lambda_k(v)$ of $\hat H(v)$ 
become asymptotic to those of $\hat H_0(v) = - \d^2 / \d q^2 + v q^M$.
We then expect ${\det}^\pm (\hat H(v) + \lambda)$ to become somehow 
asymptotically proportional to ${\det}^\pm (\hat H_0(v) + \lambda)$ 
as $v \to +\infty$, 
but the latter regime is singular and moreover non-uniform in $k$; 
hence the actual behavior of the determinants cannot be taken for granted.
In \cite{VQ} (Secs.~3--4), we tackled it for a few binomial potentials 
and exclusively at $\lambda=0$; now we will do it in full generality.
\medskip

\subsection{Detailed anomaly types}

As argued in Sec.~2.1, it is essential to distinguish between 
normal (zero-residue) and anomalous (non-zero residue) cases, 
but this now applies independently 
to the coupled ($=\hat H(v)$) and the uncoupled ($=\hat H_0(v)$) problems.

\noindent - the coupled problem ($\Pi_\lambda(q)^2 = q^N + v q^M + \lambda$):
the residue function is the coefficient of $q^{-1-Ns}$ 
in the generalized binomial expansion for 
$q^{N(1/2 \, -s)} (1+ v q^{M-N} + \lambda q^{-N})^{1/2 \, -s}$.
When $N>2$ as here, the residue function cannot involve $\lambda$; specifically,
\beq
\label{ANj}
\beta_{-1}(s) \equiv 0 \quad \mbox{unless }
{N+2 \over 2(N-M)} =j \in {\mathbb N}^\ast \quad
(\mbox{``anomaly condition ${\mathbf A}_j$ of level } j \mbox{"}) ;
\eeq
thus, anomalies attach to {\sl specially correlated exponents\/} $N,\ M$ only:
\beq
(\mbox{level } j:) \quad N=2jm-2,\ M=N-m \mbox{ for } m \in {\mathbb N}^\ast 
\quad \mbox{(with $m$ even for even potentials)} ;
\eeq
and then
\beq
\label{BTj}
\beta_ {-1}(s) \equiv (-1)^j { \G (s \!+\! j \!-\! 1/2) \over \G (s-1/2) \, j! }
\, v^j \qquad \Bigl[ \beta_ {-1}(0) = 
(-1)^{j-1} { (2j \!-\!2)! \over 2^{2j-1} (j \!-\! 1)! \, j! } \, v^j \Bigr] .
\eeq
- the uncoupled problem ($\Pi_{0,\lambda}(q)^2 = vq^M + \lambda$):
the same calculation now simply yields
\beq
\label{BT1}
\beta_ {-1}(s) \equiv v^{-1/2} \lambda \, (1/2 \, -s) \quad 
\mbox{if } M=2 \quad [{\mathbf A}_1 \mbox{ for } \lambda \ne 0 ], \qquad
\mbox{otherwise } \beta_ {-1}(s) \equiv 0 \quad [{\mathbf N}] .
\eeq
The {\sl harmonic oscillator\/} ($\Pi(q)^2 = vq^2+\lambda$)
thus gives the prime example of anomaly, actually the unique case
(among all potentials) where the residue depends on the spectral parameter;
all other binomials $\{ vq^M+\lambda \}$ ($M \ne 2$) are of type {\bf N}.

The type can abruptly change either way in the $v \to +\infty$ limit, 
giving birth to four distinct variants
(the ``basic" example of eq.(\ref{QAO}) is not the simplest!):

\noindent {\bf N} $\to$ {\bf N}: e.g., $V(q) = q^8+vq^4$;

\noindent ${\mathbf A}_j \to$ {\bf N}: e.g., $V(q) = q^6+vq^4$, of level $j=2$;

\noindent {\bf N} $\to {\mathbf A}_1$: e.g., $V(q) = q^4+vq^2$
(the ``basic" example) when $\lambda \ne 0$;

\noindent ${\mathbf A}_j \to {\mathbf A}_1$: 
only one case, $V(q) = q^6+vq^2$ when $\lambda \ne 0$, for which $j=1$.

\subsection{The main estimate}

We can relate the coupled and uncoupled spectral determinants very easily
through a key result of exact WKB theory, the basic identities (\ref{ID}).
These are to be written for both (coupled and uncoupled) problems independently:
\beq
\label{IPU}
\matrix{
\hfill {\det}^- (\hat H(v) + \lambda) \si\equiv\sf \Psi_\lambda (0,v), \hfill
{\det}^+ (\hat H(v) + \lambda) \si\equiv\sf -\Psi'_\lambda (0,v) , \cr
\hfill {\det}^- (\hat H_0(v) + \lambda) \si\equiv\sf \Psi_{0,\lambda} (0,v), 
\hfill \qquad
{\det}^+ (\hat H_0(v) + \lambda) \si\equiv\sf -\Psi'_{0,\lambda} (0,v) , }
\eeq
where $\Psi_\lambda (q,v)$, resp. $\Psi_{0,\lambda} (q,v)$ 
are the canonical recessive solutions
of $(\hat H (v)+\lambda) \Psi = 0$, resp. $(\hat H_0 (v) + \lambda) \Psi_0 = 0$.
So, the problem boils down to relating
$\Psi_\lambda (q,v)$ and $\Psi_{0,\lambda} (q,v)$ near $q=0$ as $v \to +\infty$.

Now, as soon as $|q|^{N-M} \ll v$, the term $q^N$ becomes 
a negligible perturbation of $vq^M$ within the Schr\"odinger equation, 
hence the recessive solution $\Psi_\lambda (q,v)$
has to become {\sl asymptotically proportional to\/} $\Psi_{0,\lambda} (q,v)$ 
in that regime (given that the WKB form (\ref{WKB}) holds asymptotically for
$\Pi_\lambda(q) \to +\infty$ whichever way the limit takes place, including
$v \to +\infty$ at fixed $q$). The only problem is then to determine 
the asymptotic ratio
$\Psi_\lambda (q,v) / \Psi_{0,\lambda} (q,v) \sim C(\lambda,v) $ 
($q$-independent) as $v \to +\infty$ at fixed $q$.
By contrast, the alternative normalization of recessive solutions 
{\sl based at\/} $q=0$,
\beq
\label{BKW}
{\mit \Psi}_\lambda (q,v) \sim \Pi_\lambda(q,v)^{-1/2} \e^{ -\int_0^q \Pi_\lambda(q',v) \,\d q'},
\qquad {\mit \Psi}_{0,\lambda} (q,v) \sim
\Pi_{0,\lambda}(q,v)^{-1/2} \e^{ -\int_0^q \Pi_{0,\lambda}(q',v) \,\d q'}
\eeq
(for $q \to +\infty$) immediately entails
\beq
{\mit \Psi}_\lambda (q,v) \sim {\mit \Psi}_{0,\lambda} (q,v) \qquad 
\mbox{for } v \to +\infty, \quad |q|^{N-M} \ll v ,
\eeq
because the asymptotic equivalence $\Pi_\lambda(q',v) \sim\Pi_{0,\lambda}(q',v)$
can be used all over the {\sl bounded\/} interval $[0,q]$.

The final issue is to relate the two normalizations, 
the canonical one of eq.(\ref{WKB}) (``based at $q=+\infty$") 
and the latter one based at $q=0$. 
Thanks to eq.(\ref{ADD}), the answer is simply
\beq
\label{REN}
\Psi_\lambda (q,v) \equiv 
\e^{ \int_0^{+\infty} \Pi_\lambda(q',v) \,\d q' } {\mit \Psi}_\lambda (q,v) 
\quad (\mbox{and likewise for } \Psi_{0,\lambda} \mbox{ with } \Pi_{0,\lambda}).
\eeq
Finally, putting together eqs.(\ref{IPU})--(\ref{REN}),
we end up with the comparison formula
\beq
\label{CMP}
{\det}^\pm (\hat H(v)+\lambda) \sim
\e^{S(\lambda,v)} \e^{-S_0(\lambda,v)} {\det}^\pm (\hat H_0(v)+\lambda) \qquad (v \to +\infty)
\eeq
(stated in most general terms), where
\beq
S(\lambda,v) = \int_0^{+\infty} \!\! \Pi_\lambda (q,v) \,\d q , \qquad \mbox{resp.} \quad
S_0(\lambda,v) = \int_0^{+\infty} \!\! \Pi_{0,\lambda} (q,v) \,\d q ,
\eeq
are coupled, resp. uncoupled, improper action integrals. 
Specifically here,
\beq
\label{PU}
S(\lambda,v) = \int_0^{+\infty} \!\! (q^N+vq^M+\lambda)^{1/2} \,\d q, \qquad \mbox{resp.}
\quad S_0(\lambda,v) = \int_0^{+\infty} \!\!\ (vq^M+\lambda)^{1/2} \,\d q .
\eeq
The problem has thus been decomposed and reduced
to the asymptotic ($v \to +\infty$) evaluation of 
{\sl the two action integrals\/} of eq.(\ref{PU}) separately.

\section{Explicit formulae for improper action integrals}

This Section constitutes a kind of technical digression,
but the effective computations of improper action integrals to be presented
might also be of autonomous interest and applicability.

\subsection{Binomial $\Pi(q)^2$ : exact evaluation}

We compute the improper action integral $\int_0^{+\infty} \Pi(q) \,\d q$ 
{\sl exactly\/} for a {\sl binomial\/} $ \Pi(q)^2 = uq^N+vq^M $,
in the rather general setting $N > M \ge 0$, $ u,\ v >0$,
resulting in the multi-purpose formulae (\ref{BCN}) and (\ref{BCj})
(here, $N$ and $M$ might even be non-integers).

At the core, 
$ \int_0^{+\infty} \Pi(q) \,\d q = \lim_{s=0} I_0(s) $ where
\beq
I_0(s) \defi \int_0^{+\infty} (uq^N+vq^M)^{1/2 \, -s} \,\d q 
\qquad (\Re s > \od) ,
\eeq
as long as the limit (understood as the analytical continuation to $s=0$) 
is finite. Now the right-hand side reduces to a Eulerian integral, of the form
\beq
\label{EU}
\int_0^{+\infty} (ax+b)^{1/2 \, -s} x^{\alpha-1} \d x \equiv 
a^{-\alpha} b^{1/2 \, +\alpha-s} \G (\alpha) \G (s-\alpha-1/2) / \G (s-1/2)
\eeq
(under the change of variable $q^{N-M} = u^{-1} v \, x$; 
here, $\alpha = [M(1-2s)+2] / [2(N-M)]$); more precisely,
\beq
\label{BCS}
I_0(s) \equiv {\G ( {M(1 - 2s) + 2 \over 2(N-M)} )
\G ( -{N(1 - 2s) + 2 \over 2(N-M)} ) \over (N-M) \, \G (s \!-\! 1/2) } 
\, u^{-{M(1-2s)+2 \over 2(N-M)}} v^{N(1-2s)+2 \over 2(N-M)} .
\eeq
Consequently, at $s=0$,
\beq
\label{BCN}
\int_0^{+\infty} (uq^N+vq^M)^{1/2} \,\d q = 
{ \G ( {M+2 \over 2(N-M)} ) \G ( -{N+2 \over 2(N-M)} ) \over (N-M) \G (-1/2) } 
\, u^{-{M+2 \over 2(N-M)}} v^{N+2 \over 2(N-M)} 
\eeq
{\sl in the normal case\/}, i.e., when the right-hand side is finite, meaning
here $ {\textstyle N+2 \over \textstyle 2(N \!-\! M)} \notin {\mathbb N} $.

As concrete examples of this {\bf N} type:
\bea
\label{BC4}
\int_0^{+\infty} (q^4 + v q^2)^{1/2} \,\d q \si = \sf -v^{3/2}/3 \\
\label{BCH}
\int_0^{+\infty} (u q^N + \lambda)^{1/2} \,\d q \si = \sf 
-(2 \sqrt \pi)^{-1} {\textstyle \G (1+{1 \over N})\G (-{1 \over 2}-{1 \over N})}
\, u^{-{1 \over N}} \lambda^{ {1 \over 2}+{1 \over N} }
\quad (N \ne 2).
\eea

Now, the right-hand side of eq.(\ref{BCN}) turns infinite whenever
$(2j-1)N = 2(jM+1)$ for some $j \in {\mathbb N}^\ast$ ($j=0$ cannot occur);
this is precisely the anomaly condition ${\mathbf A}_j$ of level $j$.
The binomials of any type ${\mathbf A}_j$ can be readily (albeit tediously) 
handled by applying eq.(\ref{IFP}) to $I_0(s)$. 
(The cases with $j=1$ as well as the action integral of eq.(\ref{BC4}) 
were implicitly evaluated in \cite{VQ}, by a different route.)
First, the residue is
\beq
\label{BETj}
\beta_ {-1}(0) = (-1)^{j-1} { (2j \!-\!2)! \over 2^{2j-1} (j \!-\! 1)! \, j! }
\, u^{1/2 \, -j} v^j ;
\eeq
then, the finite part at $s=0$ of eq.(\ref{BCS}) gets extracted as
\[
\beta_{-1}(0) \Bigl( {2j \over N \!+\! 2} 
\Bigl[ \psi(j+1) - \log v + {M \over N} \Bigl( -\psi(j-1/2)+\log u \Bigr) \Bigr]
- {1 \over N} \, \psi(-1/2) \Bigr) ;
\]
so that finally, when $M = [(j-1/2)N-1]/j$ (condition ${\mathbf A}_j$),
eq.(\ref{IFP}) yields
\beq
\label{BCj}
\int_0^{+\infty} \!\! (uq^N \!+\! vq^M)^{1/2} \,\d q =
{ 2j \, \beta_{-1}(0) \over N+2 } \Bigl[ -\log v + \sum_{m=1}^j {1 \over m} 
+ {2M \over N} \Bigl( 
\log 2 + \hf \log u - \sum_{m=1}^{j-1} {1 \over 2m \!-\! 1} \Bigr) \Bigr] .
\eeq

The cases with $j=1$ are of special interest.
Besides the harmonic oscillator, the general binomials of type ${\mathbf A}_1$
are just the {\sl supersymmetric potentials\/} (at zero energy):
\beq
\Pi(q)^2 = uq^N+vq^M \qquad \mbox{with } N=2M+2 \quad (M>0) ,
\eeq
and eq.(\ref{BCj}) distinctly simplifies to
\beq
\label{BC1}
\int_0^{+\infty} \!\! (uq^N \!+\! vq^{N/2 \, -1})^{1/2} \,\d q = 
{u^{-1/2} v \over N \!+\! 2} 
\Bigl[-\log v + 1 + {N \!-\! 2 \over N} (\log 2 + \hf \log u ) \Bigr] 
\quad (j=1).
\eeq
In particular, for the harmonic oscillator ($N=2$)
at a general energy value $(-\lambda)$,
\beq
\label{BC2}
\int_0^{+\infty} (vq^2 + \lambda)^{1/2} \,\d q = 
\qt v^{-1/2} \lambda (1 - \log \lambda).
\eeq

\subsection{Trinomial $ \Pi(q)^2 $ : asymptotic $v \to \infty$ evaluation}

We now consider a trinomial $ \Pi(q)^2 $ of the form $ q^N+vq^M+\lambda $,
with even $N>M>0$, and a systematically {\sl constant\/} third term:
the spectral parameter itself, $\lambda \ (>0)$ (= minus the total energy).
One of the coefficients can always be scaled out to unity, 
and we have done this for the highest power initially.

In this case we can no longer compute the action integral
$\int_0^{+\infty} \Pi(q) \,\d q$ exactly.
In view of eq.(\ref{CMP}), however, we mainly need its large-$v$ behavior,
specially for $v \to +\infty$ in order to recover
singular perturbation theory according to eq.(\ref{SYZ}) 
(but as in \cite{VQ}, we expect the results to remain valid 
over suitable sectors in the complex $v$-plane). 

According to the zeta-regularization idea, 
we must start from the large-$v$ behavior of 
$I(s;\lambda,v) \defi \int_0^{+\infty} (q^N+vq^M+\lambda)^{1/2 \, - s} \, \d q$;
this problem is rather delicate, so any brute-force expansion scheme is dubious.
Instead, we apply the following general idea: 
if the function $I(v)$ under study is an inverse Mellin transform, 
\beq
\label{IMT}
I(v) = (2\pi\mi)^{-1} \int_{c-\mi \infty}^{c+\mi \infty} \tilde I(\sigma) v^\sigma \,\d \sigma,
\eeq
then the singularities of $\tilde I(\sigma)$ in the half-plane 
$\{ \Re \sigma < c \}$ encode the large-$v$ behavior of $I(v)$. 
Thus (by the residue calculus) any polar part of the form 
$A(\sigma - \sigma_0)^{-2} + B(\sigma - \sigma_0)^{-1}$ in $\tilde I(\sigma)$
represents an asymptotic contribution $v^{\sigma_0} (A \log v + B)$ to $I(v)$.
This is particularly valuable for $I(s;\lambda,v)$, because its direct Mellin transform 
$\tilde I(s;\lambda,\sigma) \defi \int_0^{+\infty} I(s;\lambda,v) \,
v^{-\sigma-1} \,\d v$ is {\sl exactly computable\/}
(using the same formula (\ref{EU}) as for the exact action integral 
of a binomial but now {\sl twice in succession\/}), 
and it is a meromorphic function of $\sigma$: formally,
\[
\int_0^{+\infty} \!\!\!\d v \, v^{-\sigma-1} 
(q^N \!+\! vq^M \!+\! \lambda)^{1/2 \, - s} =
{\G(-\sigma) \G (s \!+\! \sigma \!-\! 1/2) \over \G (s -1/2) }
\, q^{M\sigma}(q^N \!+\! \lambda)^{1/2\, -s-\sigma}
\]
\beq
\label{MEL}
\Longrightarrow \quad
\tilde I(s;\lambda,\sigma) = {\G(-\sigma) \G ({M \sigma +1 \over N}) 
\G (s \!+\! \sigma \!-\! \hf \!-\! {M \sigma + 1 \over N}) \over N \G (s -1/2) }
\, \lambda^{ -s - {N-M \over N }\sigma + {1 \over 2}+ {1 \over N} } 
\eeq
(using the change of variable $q^N=\lambda \, r$ for the $q$-integration).
Now this Mellin transform also has to genuinely exist somewhere;
here, all integrations converge in some strip 
$\sigma' < \Re \sigma <0$ provided $\Re s > \mu(N)$, 
and the inverse transformation (\ref{IMT}) applies with $c=-0$.
Consequently, the poles $\sigma(s)$ relevant to the current asymptotic problem
are those which lie in $\{ \Re \sigma <0 \}$ when $\Re s > \mu(N)$, 
and their contributions are then to be analytically continued to $s=0$.
Overall, the poles in eq.(\ref{MEL}) form three arithmetic progressions,
one for each Gamma factor in numerator; they are real for real $s$ (fig.~1).
At $s=0$, any pole $\sigma(s)$ will contribute an asymptotic term 
of degree $d_v=\sigma(0)$ in $v$ (on general grounds)
and of degree $d_\lambda={1 \over 2}+ {1 \over N} - {N-M \over N }\sigma(0)$ 
in $\lambda$ (by examination of eq.(\ref{MEL})).
At the end, we plan to keep the terms of degree $d_g \le 0$ 
in the perturbative coupling constant $g$ 
(discarding $o(1)$ terms when $g \to 0$);
now the Symanzik scaling (eq.(\ref{NMO}) at fixed $E$) entails $d_g \equiv 
-{M+2 \over N+2} (d_v + {2 \over M+2} \, d_\lambda) = -(M \sigma(0) +1)/N$;
altogether, $d_g \le 0$ then amounts to keeping only the poles for which $\sigma(0) \ge -1/M$.

\begin{figure}
{\hfill
\epsfysize=6cm
\epsfbox{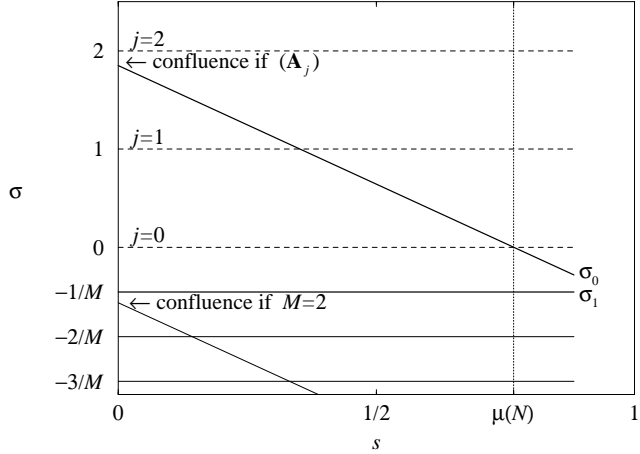}
\hfill}
\caption{\small Schematic plot (using some non-integer $N$, $M$)
of the poles $\sigma(s)$ of the Mellin transform $\tilde I(s;\lambda,\sigma)$ 
in eq.(\ref{MEL}). The two main contributing poles 
(in the $v \to +\infty$ limit) are drawn with bold lines; 
non-contributing poles are drawn with dashed lines.
}
\end{figure}

When $M \ge 2$ (as here), only two poles $\sigma (s)$ satisfy both criteria, 
(in real form) $\sigma(s)<0$ for $s>\mu(N)$ and $\sigma(0) \ge -1/M$: 
they are, in decreasing order (at $s=0$),
\beq
\label{POL}
\sigma_0(s) \equiv {N \over N-M} \Bigl( {1 \over 2} + {1 \over N}-s \Bigr) 
\quad \mbox{(leading) \qquad and} \quad 
\sigma_1(s) \equiv -{1 \over M} \quad \mbox{(subleading)} .
\eeq
They are generically simple, with two exceptions at $s=0$:
$\sigma_0(0) = {N+2 \over 2(N-M)}$
becomes confluent with the (fixed) pole $\sigma=+j$
when the {\sl coupled\/} problem is of anomalous type ${\mathbf A}_j$;
and independently, $\sigma_1$ becomes confluent with the (next mobile) pole 
${N \over N-M} (\hf + {1 \over N}-s-1)$ 
when the {\sl uncoupled\/} problem is anomalous, i.e., $M=2$.
The latter will induce a usual double-pole contribution; 
the former confluence is worse, \
making the inverse-Mellin representation {\sl singular\/} 
as the integration path gets {\sl pinched\/} between the two confluent poles.

We now specifically evaluate the two dominant polar contributions, 
from $\sigma_0$ and $\sigma_1$.

\noindent - the leading pole $\sigma_0 (0) = {N+2 \over 2(N-M)}$:
if the coupled problem is of type {\bf N} this pole remains simple, 
and its asymptotic contribution 
$[\Res_{\sigma_0} \tilde I(s;\lambda,\sigma)] \, v^{\sigma_0}$
turns out (by inspection) to be just $\int_0^{+\infty} (q^N+vq^M)^{1/2} \,\d q$
(as given by eq.(\ref{BCN}) at $u=1$). 
Furthermore, 
$ \partial_\lambda I(s;\lambda,\sigma) \propto I(s+1;\lambda,\sigma) $,
an operation which precisely annihilates this leading pole part in all cases, 
so the latter has to be a constant in $\lambda$; 
then, its computation at $\lambda=0$ precisely yields 
$\int_0^{+\infty} (q^N+vq^M)^{1/2} \,\d q$, 
{\sl now including the confluent cases\/} (${\mathbf A}_j$).

\noindent - the subleading pole $\sigma_1 = -1/M$:
if $M>2$ this pole remains simple, and its asymptotic contribution 
$[\Res_{\sigma_1} \tilde I(s;\lambda,\sigma)] \, v^{\sigma_1}$
coincides with $\int_0^{+\infty} (vq^M+\lambda)^{1/2} \,\d q$ 
as given by eq.(\ref{BCH}).
Under confluence ($M=2$), 
the contribution becomes that of the double pole of eq.(\ref{MEL}) 
at $\sigma=-1/2$: this is 
$\int_0^{+\infty} (vq^2+\lambda)^{1/2} \,\d q + {N \over N-2} A_1(\lambda,v)$
where the action integral is given by eq.(\ref{BC2}), and 
\beq
\label{AN2}
A_1(\lambda,v) = \qt v^{-1/2}\lambda (\log v + 2 \log 2) .
\eeq

All in all, the asymptotic $v \to + \infty$ formula for the trinomial action integral is then
\beq
\label{AST}
\int_0^{+\infty} \!\! (q^N+vq^M+\lambda)^{1/2} \, \d q \sim
\!\int_0^{+\infty} \!\! (q^N+vq^M)^{1/2} \,\d q + \!\int_0^{+\infty} \!\! (vq^M+\lambda)^{1/2} \,\d q
+ \delta_{M,2} \, {N \over N \!-\! 2} \, A_1(\lambda,v) ,
\eeq
where $\int_0^{+\infty} (q^N+vq^M)^{1/2} \, \d q$ is specified 
through eq.(\ref{BCN}) if the coupled problem is of type {\bf N}, 
or else eq.(\ref{BCj}) if the coupled problem is of type ${\mathbf A}_j$ 
(i.e., if ${N+2 \over 2(N-M)} = j \in {\mathbb N}^\ast$);
whereas $\int_0^{+\infty} (vq^M+\lambda)^{1/2} \,\d q$ is given by 
eq.(\ref{BCH}) if $M>2$, or eq.(\ref{BC2}) if $M=2$ 
(and $\delta_{M,2}$ is a Kronecker delta symbol).

\section{Asymptotic behaviors of spectral functions}

The theoretical results of Secs.~3--4 translate into concrete formulae 
for spectral functions in the $v \to +\infty$ regime.

\subsection{The spectral determinants}

Upon substituting the explicit formulae of Sec.~4 into eq.(\ref{CMP}), 
$\int_0^{+\infty} (vq^M+\lambda)^{1/2} \,\d q$ cancels out, 
and a slightly simpler $v \to +\infty$ formula results:
\beq
\label{AS}
{\det}^\pm (- \d^2/ \d q^2 + q^N + vq^M + \lambda) \sim 
\e^{\int_0^{+\infty} (q^N+vq^M)^{1/2} \d q
+ \delta_{M,2} \,{N \over N-2} A_1(\lambda,v)} 
{\det}^\pm (- \d^2/ \d q^2 + vq^M+\lambda),
\eeq
where $\int_0^{+\infty} (q^N+vq^M)^{1/2} \,\d q$ is given through eq.(\ref{BCN})
if the coupled problem is of type {\bf N}, or (\ref{BCj}) if it is of type 
${\mathbf A}_j$, and $A_1(\lambda,v)$ by eq.(\ref{AN2}).

Being homogeneous, 
the uncoupled potentials obey a simpler form of the scaling eq.(\ref{SYZ}): 
$ (- \d^2/ \d q^2 + vq^M)$ is unitarily equivalent to 
$v^{2 /(M+2)} (- \d^2/ \d q^2 + q^M)$;
then the scaling laws (\ref{SC}) apply with $r=v^{2 /(M+2)}$;
as these laws are more awkward for the ${\det}^\pm$ 
than for the full and skew determinants $\det$ and ${\det}^{\rm P}$, 
we now switch to the latter combinations and explicitly get
\bea
\label{HS}
\det (- \d^2/ \d q^2 + vq^M + \lambda) \si \equiv \sf 
\det (- \d^2/ \d q^2 + q^M + v^{-2 /(M+2)} \lambda ) \qquad \quad(M \ne 2) \\
\label{HOS}
\det (- \d^2/ \d q^2 + vq^2 + \lambda) \si \equiv \sf 
v^{-v^{-1/2}\lambda /4} \det (- \d^2/ \d q^2 + q^2 + v^{-1/2}\lambda ) \\
\label{FF}
{\det}^{\rm P} (- \d^2/ \d q^2 + vq^M + \lambda) \si \equiv \sf v^{1 /(M+2)} 
\, {\det}^{\rm P} (- \d^2/ \d q^2 + q^M + v^{-2 /(M+2)} \lambda ) .
\eea
Remark: the harmonic-oscillator determinants are actually known in closed form 
(\cite{VQ}, eqs.(155));
e.g., eq.(\ref{HOS}) has the fully explicit form (to be used later)
\beq
\label{HO}
\det (- \d^2/ \d q^2 + vq^2 + \lambda) \equiv 
v^{-v^{-1/2}\lambda /4} 2^{-v^{-1/2}\lambda/2} \sqrt{2\pi} / \G(\hf (1+v^{-1/2}\lambda )) .
\eeq
Thus, eqs.(\ref{AS}) plus (\ref{HS})--(\ref{FF}) supply 
the $v \to +\infty$ behaviors at fixed $\lambda$ of the coupled determinants 
in terms of the corresponding uncoupled determinants at $\lambda=0$ 
(which are computable numbers, cf. \cite{VQ}, eq.(136)).
\medskip

However, our main concern is rather the singular perturbation limit:
$v \equiv g^{-(M+2)/(N+2)} \to +\infty$ with $
v^{-2 /(M+2)} \lambda \defi (-E)$ fixed, 
according to eq.(\ref{NMO}). The explicit final results,
deduced from eqs.(\ref{AS})--(\ref{FF}) after rescaling both sides, are then
\bea
\label{ASF}
\det (- \d^2/ \d q^2 \si + \sf q^M + g q^N - E) \, /
\det (- \d^2/ \d q^2 + q^M - E) \nonumber \\
\si \sim \sf \e^{2 \int_0^{+\infty} (q^N+vq^M)^{1/2} \d q}
\e^{- \delta_{M,2} \,{1 \over N-2} ({N+2 \over 4} \log v + N \log 2) E} 
\qquad \qquad \! \mbox{for type \bf N} \\
\si \sim \sf v^{-{4 \beta_{-1}(0) \over N(M+2)} }
\e^{2 \int_0^{+\infty} (q^N+vq^M)^{1/2} \d q} 
\e^{- \delta_{M,2} \,{1 \over N-2} ({N+2 \over 4} \log v + N \log 2) E} 
\ \mbox{for type } {\mathbf A}_j \nonumber
\eea
($\beta_{-1}(0)$ (given by eq.(\ref{BETj})), and type, 
both refer to the coupled problem);
whereas the skew determinants always behave straightforwardly:
\beq
{\det}^{\rm P} (- \d^2/ \d q^2 + q^M + g q^N - E) \sim 
{\det}^{\rm P} (- \d^2/ \d q^2 + q^M - E) .
\eeq
The main result here is the explicit non-trivial prefactor in eq.(\ref{ASF}).
Its essential singularity for $g \to 0$ should relate to 
the non-uniformity of this limit with respect to the quantum number $k$.
By contrast, the dependence of its logarithm upon $E$ is elementary,
consisting only of (a) {\sl constant\/} term(s) 
(already determined in \cite{VQ} for some cases),
then a {\sl linear\/} term, and nothing else.
The basic example (\ref{QAO}), being of type {\bf N}, thus gives
\beq
\det (- \d^2/ \d q^2 + q^2 + g q^4 - E) \sim 
\e^{-2/ \, 3g} \e^{(\log g \, /2 \, - 2 \log 2) E} 
\det (- \d^2/ \d q^2 + q^2 - E) .
\eeq
(Note the ``instanton-like" structure of the first prefactor, 
computed by eq.(\ref{BC4}).)

\subsection{The spectral zeta functions}

Over the spectrum $\{ E_k(g) \}$ of the rescaled operator 
$(- \d^2/ \d q^2 + q^M + g q^N)$, we can consider the full and skew
spectral zeta functions
\beq
\label{ZPD}
Z_g(s;E) \defi \sum_{k=0}^\infty (E_k(g)-E)^{-s}, \qquad 
Z_g^{\rm P}(s;E) \defi \sum_{k=0}^\infty (-1)^k (E_k(g)-E)^{-s}
\eeq
for, say, integer $s \in{\mathbb N}^\ast$, in which case they converge for $g>0$
and relate to the spectral determinants in a simpler way than for general $s$,
\beq
\label{ZS}
Z_g(s;E) \equiv -{1 \over (s-1)!} { \partial^s \over \partial E^s} 
\log \, \det (- \d^2/ \d q^2 + q^M + g q^N - E) ,
\eeq
(obtained from eq.(\ref{WPR}) upon rescaling; 
and likewise for $(Z^{\rm P},\ {\det}^{\rm P}$)).

\begin{figure}
{\hfill
\epsfysize=6.5cm
\epsfbox{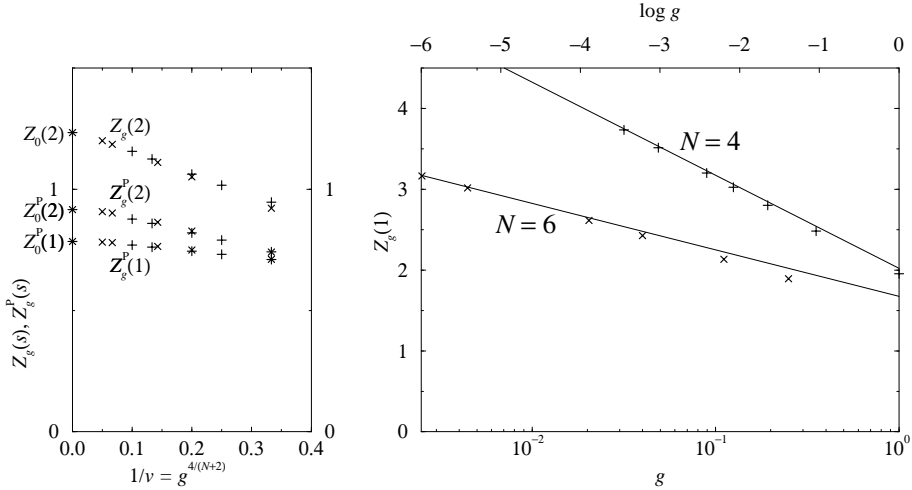}
\hfill}
\caption{\small Illustration of the two $g \to 0$ behaviors 
of spectral zeta functions at $E=0$, $Z_g(s) \defi Z_g(s;0)$ and 
$Z_g^{\rm P}(s) \defi Z_g^{\rm P}(s;0)$, as computed from numerical spectra 
(by means of eqs.(\ref{ZPD})) for a sample of potentials $V(q)=q^2+ g q^N$ 
with $N=4$ ({\bf +}) or $6$ ({\boldmath $\times$}).
Left: regular behaviors verifying eq.(\ref{ASR}), plotted against $1/v$
(the variable which empirically appears to yield the simplest dependence);
remark: $Z_0^{\rm P}(1)=\pi/4$, $Z_0(2)=\pi^2/8$, 
and $Z_0^{\rm P}(2)= \mbox{\sl Catalan's constant \/} (\approx 0.9159656)$.
Right: singular behavior of $Z_g(1)$, plotted in a $\log g$ coordinate;
the straight lines show the theoretical asymptotic predictions (eq.(\ref{RF})).
}
\end{figure}

Assuming all previous estimates are stable under $E$-differentiations 
(as is usually the case in WKB theory), the preceding formulae imply
the {\sl regular behaviors\/} (see fig.~2, left)
\beq
\label{ASR}
Z_g(s;E) \sim Z_0(s;E), \quad 
Z_g^{\rm P}(s;E) \sim Z_0^{\rm P}(s;E) \qquad \quad (g \to 0) ,
\eeq
{\sl except for\/} $Z_g(1;E)$ (the resolvent trace) {\sl when\/} $M=2$, 
which gives {\sl the singular case\/} 
($Z_0(1;E)$ infinite, while $Z_0^{\rm P}(1;E)$ stays finite).
Those patterns were conjectured in \cite{VQ} (Sec.~3), 
but not the precise divergent behavior of $Z_g(1;E)$, which requires to know
the $E$-linear term in the exponent of the determinant ratio (\ref{ASF}).
For $s=1$, eq.(\ref{ZS}) needs to be regularized at $g=0$, as
\beq
- ( \d / \d E) \log \det (- \d^2/ \d q^2 + q^2-E)
\equiv -\hf [ \psi (\hf (1-E)) + \log 2 ] 
\eeq
(using the known closed form (\ref{HO}) of the harmonic-oscillator determinant).
Then the logarithmic differentiation of eqs.(\ref{ASF}) for $M=2$ 
yields the $g \to 0$ behavior of $Z_g(1;E)$ for all potentials $q^2+gq^N$ 
(irrespective of type), as the following singular expression:
\beq
Z_g(1;E) \sim 
{1 \over N-2} (- \log g + N \log 2) - \hf \, [ \psi (\hf (1-E)) + \log 2 ] .
\eeq
For instance, at $E=0$ this gives (see fig.~2, right)
\bea
\label{RF}
\sum_{k=0}^\infty E_k(g)^{-1} \si \sim \sf 
-{1 \over N-2} \log g + {1 \over 2} \Bigl( \g + {3N-2 \over N-2}\log 2 \Bigr) 
\qquad \qquad (g \to 0) \\
\si \sim \sf - \hf \log g + \hf (\g + 5 \log 2) \ \mbox{ for } N=4,
\quad - \qt \log g + \hf \g + 2 \log 2 \ \mbox{ for } N=6, \ldots \nonumber
\eea
(to be compared with the sharp cutoff regularization of eq.(\ref{GM})).

\subsection{Concluding remarks}

We have completed here one ``exercise in exact quantization" begun in \cite{VQ}:
we gave the $g\to 0$ behavior of the spectral determinants
${\det}^\pm (- \d^2/ \d q^2 + q^M + g q^N - E)$, now for general parameter values.
While it may appear wasteful to use a wholly exact approach 
for perturbative calculations,
exact WKB analysis actually proved quite efficient for the task;
inversely, such problems help to strengthen the practical sides of that field, 
which still need further development.

We are also confident that the above approach can be extended further,
both to complex parameter asymptotics 
and towards higher orders in powers of $g$.

\end{document}